\documentclass[preprint,3p,times,12pt]{elsarticle}

\usepackage{amssymb}
\usepackage{amsmath}

\usepackage{graphicx}
\usepackage{dcolumn}
\usepackage{booktabs}
\usepackage{siunitx}
\usepackage[colorlinks=false,linktocpage=true]{hyperref}
\usepackage{setspace}
\usepackage{multibib}
\usepackage{array}
\usepackage{tabularx}
\usepackage{algorithm}
\usepackage{algpseudocode}
\usepackage{subcaption}
\usepackage{multirow}
\usepackage{avant}
\usepackage{cmbright}
\usepackage{siunitx}
\usepackage{pgfplotstable}
\usepackage{booktabs}
\usepackage{csvsimple}
\pgfplotsset{compat=1.15}
\usepackage{makecell}
\usepackage{cmbright}

\usepackage{accents}

\newcommand{\T}[1]{\boldsymbol{#1}}
\newcommand{\TT}[1]{\boldsymbol{#1}}

\newcommand{\average}[1]{\left<#1\right>}

\journal{PEP}

\begin{document}

\begin{frontmatter}

\title{Elastic anisotropy of 1,3,5-Triamino-2,4,6-Trinitrobenzene as a function of temperature and pressure: A Molecular Dynamics study}

\author[1,2]{Paul Lafourcade\corref{cor1}}
\ead{paul.lafourcade@cea.fr}
\author[1,2]{Jean-Bernard Maillet}
\author[1,2]{Nicolas Bruzy}
\author[1,2]{Christophe Denoual}

\address[1]{CEA DAM DIF, 91297 Arpajon CEDEX, France}
\address[2]{LCME, Universit\'e Paris-Saclay, France}
\cortext[cor1]{Corresponding author}

\date{\today}

\begin{abstract}
The equation of state of the triclinic compound 1,3,5-triamino-2,4,6-trinitrobenzene (TATB) as well as its second-order isothermal elastic tensor were computed through classical molecular dynamics simulations under various temperature and pressure conditions. Hydrostatic pressures similar to previous diamond anvil cell experiments were imposed within the range [0,66] GPa and temperatures chosen between 100 and 900 K in conjunction with the most recent version of an all-atom fully-flexible molecule force field. The isothermal elastic constants were computed using the generalized Hooke’s law by fitting Cauchy stress vs. linear strain curves. Along isobaric pathways, TATB single crystal stiffness is found to undergo linear softening, less pronounced at high pressure, while maintaining its elastic anisotropy. On the other hand, along an isothermal pathways, a non-linear increase is observed in the elastic constants with a significant decrease in anisotropy. Towards a precise mesoscopic modeling of TATB single crystal mechanical behavior, we provide “ready to plug-in” analytical formulations of the P,V,T equation of state and pressure-temperature dependent non-linear elasticity.
\end{abstract}

\end{frontmatter}


\section{Introduction}
\label{sec:introduction}
1,3,5-Triamino-2,4,6-trinitrobenzene (TATB) is a highly insensitive energetic molecular crystal used in multiple explosive formulations. It crystallizes in a triclinic cell of space group $P\bar{1}$ \cite{cady_crystal_1965} and its layered structure leads to very anisotropic thermo-mechanical and chemical behavior. These anisotropic properties are the consequences of the contrast between interatomic interactions within the molecular crystal, like hydrogen bonds between nitro and amino groups and van der Walls interactions across the molecular layers. Various studies on the anisotropic behavior of TATB emerged in the last few years and a strong coupling has been found between multiple deformation mechanisms such as elastic buckling, twinning and crystal plasticity \cite{lafourcade_irreversible_2018}, which led to a first mesoscopic model in which non-linear elastic and transformational behavior are described in a large deformation, Lagrangian formalism \cite{lafourcade_mesoscopic_2019}. \\

\noindent For low-symmetry molecular crystals such as TATB and other energetic materials, it has been very difficult to measure the elastic constants in an accurate way. Very few experiments have been conducted to characterize the mechanical behavior of TATB \cite{taw_mechanical_2017, steele_pressure-induced_2019}. Various EOS were measured with a diamond anvil cell (DAC) using powder diffraction (XRD) patterns \cite{olinger_hydrostatic_1976, stevens_hydrostatic_2008, plisson_equation_2017, bedrov_molecular_2009}, under hydrostatic pressure conditions, even if different types of sample were considered, i.e. powder vs small single crystals aggregates. Only one study reported a structural phase transformation of TATB \cite{steele_high-pressure_2020} leading to a layered structure, similar to $P\bar{1}$ but with a basal-plane sliding rearrangement of molecular layers. \\

\noindent In the last decade, molecular dynamics (MD) simulations proved to be a good alternative to fill the gap between experimental results and what is needed to build a reliable mesoscopic model that would allow characterizing the TATB mechanical behavior as a function of temperature and pressure, outside the 300 K isotherm generally considered in experiments. TATB EOS have been computed through classical MD simulations and ab-initio in the past \cite{bedrov_molecular_2009, rykounov_investigation_2015, fan_theoretical_2017, lafourcade_dislocation_2017, qin_first-principles_2019, mathew_generalized_2015} and were in good agreement with experiments. Elastic constants have also been computed \cite{bedrov_molecular_2009, valenzano_accurate_2012, mathew_generalized_2015, rykounov_investigation_2015, fan_theoretical_2017, lafourcade_dislocation_2017, qin_first-principles_2019} over restrained ranges of temperature and pressure, but could not be verified against experimental results, not available at this time. Various numerical studies focused on TATB thermal and chemical properties \cite{kroonblawd_theoretical_2013, kroonblawd_theoretical_2014, mathew_predicted_2018}, as well as on the identification and characterization of its deformation mechanisms \cite{lafourcade_dislocation_2017, lafourcade_detection_2018, lafourcade_irreversible_2018, lafourcade_mesoscopic_2019, lafourcade_multiscale_2018, mathew_generalized_2015, mathew_nanoindentation_2016}. A precise review and very detailed analysis of the literature toward mesoscopic modeling of high explosives has been recently published \cite{handley_calibrating_2017}.\\

\noindent In the present study, we focus on the temperature and pressure dependence of TATB crystal structure and elastic properties, which constitute the foundation for future mesoscopic modeling at the grain-scale and above, aiming at studying the mechanical behavior of polycrystalline samples under extreme conditions such as shock loading, high strain-rate deformation or thermal aggression \cite{lafourcade_mesoscopic_2019, bennett_thermo-elastoplastic_2020, gasnier_thermoelastic_2018-1, gasnier_thermoelastic_2018,  ambos_numerical_2015, gasnier_fourier-based_2015, trumel_irreversible_2021}. Computing crystal structure and elastic constants of non-reacted and crystalline TATB over a wider range of temperature and pressure is then of primary importance. The ZND (Zeldovich, von Neuman, Döring) pressure spike has been estimated somewhere between 33 and 43 GPa \cite{sollier_novel_2016, tarver_detonation_1997, sollier_dynamics_2012}. This shows the necessity to characterize the unreacted material properties below such pressure if one wants to build a mesoscopic modeling that includes a coupling between a thermo-mechanical constitutive law and a thermochemistry model. Concerning the temperature range, we considered temperatures up to 900 K, which corresponds to the melting point at ambient pressure, located between 700-900 K \cite{mathew_predicted_2018,  stolovy_exothermic_1983, gibbs_et_al_editors_2020}.\\

\noindent The paper is organized as follow. The first section is dedicated to the methods used to compute the equation of state of TATB single crystal and the second-order elastic tensor, as well as a short presentation of the force field used in the MD simulations. In the second section, the results concerning the EOS and the elastic constants are presented. Finally, the last section aims at providing “ready to plug-in” analytical formulations that can be included in a grain-scale mesoscopic model of TATB, with non-linear elasticity and volume ratio as a function of temperature and pressure.

\section{Methods}
\subsection{Force field}

The LAMMPS code was used to perform all the classical molecular dynamics (MD) simulations with a modified version of the widely used non-reactive force field (FF) for TATB initially developed by \cite{bedrov_molecular_2009} with fully-flexible molecules and 3D-periodic boundary conditions. This version of the TATB FF includes two main modifications; a repulsive intramolecular potential for OH bonds as well as tuned harmonic bond and angle potentials \cite{mathew_generalized_2015, kroonblawd_theoretical_2013}. Non-bonded interactions were computed with a cutoff distance of 11 angstrom. The Wolf potential \cite{wolf_exact_1999} was used to evaluate the electrostatic contributions with an 11 angstrom cutoff. An extensive description of this FF can be found in recent publications ~\cite{zhao_tandem_2020, zhao_strongly_2021}. Trajectories in the isothermal (NVT) ensemble were integrated using a Nosé-Hoover style thermostat \cite{nose_unified_1984, hoover_canonical_1985} with a 0.1 ps coupling constant. A barostat with a 1 ps damping constant was used for the isothermal-isobaric (NPT) ensemble \cite{martyna_constant_1994}. MD trajectories are integrated using a 0.5 fs timestep.\\

\noindent This force field and its modified versions have been used in many studies related to the thermo-mechanical behavior of TATB single crystal  under high strain rate loading \cite{lafourcade_dislocation_2017, lafourcade_irreversible_2018, lafourcade_mesoscopic_2019}, shock compression \cite{zhao_tandem_2020, zhao_strongly_2021} and also to compute its basic thermal and elastic properties \cite{bedrov_molecular_2009, kroonblawd_theoretical_2013, kroonblawd_theoretical_2014, mathew_generalized_2015, lafourcade_dislocation_2017, mathew_predicted_2018}.

\subsection{Lattice equilibration at finite temperature and pressure}
In order to obtain the lattice parameters of TATB single crystal for various temperature and pressure, a simulation cell consisting in a 8x8x12 replica of the $P\bar{1}$ structure \cite{cady_crystal_1965} was initially considered, containing 36 864 atoms, or 1536 molecules. For each (T, P) condition, a triclinic barostat with independent coupling for each cell parameters was used, allowing the triclinic cell to minimize the deviatoric part of the Cauchy stress tensor. A 3-step procedure is followed:

\begin{itemize}
    \item[-] For each temperature the system is equilibrated in the NVT ensemble during 500 ps. At the end of this trajectory, the stress tensor is not systematically hydrostatic.
    \item[-] Starting from the end of the previous NVT simulations, a NPT ramp at a 5 GPa/ns rate is performed in order to reach the various target pressures. This step helps to remove the deviatoric stress, thus leading to a fully hydrostatic stress tensor, but with potentially different lattice parameters than the ones obtained by \cite{plisson_equation_2017}.
    \item[-] Finally, NPT simulations are performed for 1 ns and the lattice parameters are averaged over the last 500 ps, during which the stress tensor is hydrostatic and no evolution is observed in the lattice lengths and angles. From the averaged lattice parameters, the compression ratio $V/V_0$ is computed, taking the ambient conditions (300 K, 0 GPa) as the reference conditions.
\end{itemize}

\noindent The thermodynamic path described above ensures a continuity of the volume evolution as a function of temperature and pressure. It was used to obtain the lattice parameters for 18 pressures between 0 and 61.8 GPa, each at 100, 300, 500, 700 and 900 K, leading to a total set of 90 equilibrated crystal structures. \\

\noindent However, for ambient pressure simulations, the NPT simulations led to isolated stacking fault which modified the out-of-plane angles and c-length, leading to a discontinuity in the EOS surface. This was overcome by unloading to $0 GPa$ the defect-free cells with hydrostatic pressure of $2 GPa$, allowing to obtain a defect-free simulation cell at ambient pressure.

\subsection{Elastic constants computation}
For any (T, P) state, the following procedure was used to compute the isothermal elastic response of TATB single crystal to an applied strain:
\begin{itemize}
    \item A 3D-periodic crystal consisting in a 8x8x12 replica of TATB unit-cell is equilibrated for 250 ps in the NVT ensemble, using the cell parameters obtained through the procedure explained in section 2.2. for each (T, P) state considered. We verified that the computed stress tensor calculated as the volume average of the virial \cite{tadmor_continuum_2011} along this trajectory was fully hydrostatic and met the expected pressure.
    \item Then, for a prescribed strain component, an NVT simulation is performed for 4 ns. The stress tensor is averaged over the last 3 ns for the equilibrated strained system. For each strain, an equilibrium stress tensor is then obtained. The slope of the stress-strain curve gives one component of the isothermal spatial elasticity tensor.
\end{itemize}

\begin{figure}
    \centering
    \includegraphics[height=8cm]{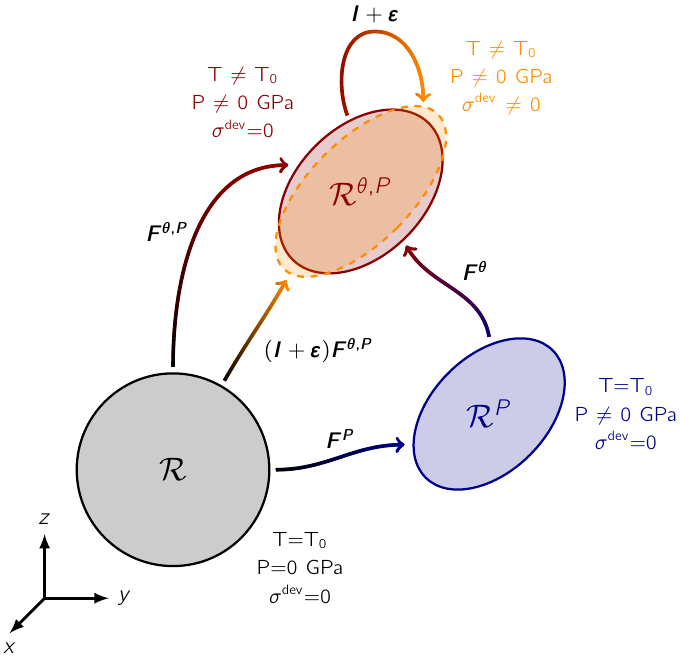}
    \caption{Multiplicative decomposition of the total deformation induced by pressure, temperature and small strain used to compute the spatial elasticity tensor of TATB single crystal. $\T{F^P}$ and $\T{F^\theta}$ respectively stand for hydrostatic pressure-induced and temperature induced deformation w.r.t reference configuration $\mathcal{R}$. $\T{F^{\theta,P}}=\T{F^\theta}\T{F^P}$ is the combination of pressure then temperature induced transformations with intermediate and final configurations $\mathcal{R}^P$ and $\mathcal{R}^{\theta,P}$. The final sate $\mathcal{R}^{\theta,P}$ is the one to which is applied the small strain tensor $\T{\varepsilon}$.}
    \label{fig:fig1}
\end{figure}

\noindent TATB crystal structure is triclinic and the single crystal has 21 independent elastic constants, which is the most general case for elasticity. Taking (300 K, 0 GPa) as the reference state for our elastic constants calculation, we define the basis transformation tensor as:
\begin{equation}
    \T{H^0} = \begin{pmatrix} a_x & b_x & c_x \\ 0 & b_y & c_y \\ 0 & 0 & c_z \end{pmatrix}
\end{equation}
where a, b and c correspond to the TATB triclinic unit cell vertices at $(T=300K, P=0GPa)$. When the material is subjected to a hydrostatic pressure and/or temperature variation, TATB unit cell undergoes a deformation that can be non-symmetric. The new basis transformation tensor such hydrostatic pressure and imposed temperature $\T{H^{\theta,P}}$ reads:
\begin{equation}
    \T{H^{\theta,P}} = \T{F^{\theta,P}} \T{H^0}
\end{equation}
where $\T{F^{\theta,P}}$ is inducing a volume change between the volume $V_0$ at (300 K, 0 GPa). This strain is a diagonal tensor in the case of an isotropic material but, in the present case, $\T{F^{\theta,P}}$ can take any form, allowing the lengths and angles of TATB unit cell to vary independently from each other. Finally, a small strain $\T{F}^\varepsilon = \T{I} + \T{\varepsilon}$ (with $\T{\varepsilon}$ a symmetric tensor) is applied on the hydrostatic pressure state. We define the new basis transformation tensor as:
\begin{equation}
    \T{H^{\theta,P,\varepsilon}} = (\T{I}+\T{\varepsilon}) \T{H^{\theta,P}} = (\T{I}+\T{\varepsilon}) \T{F^{\theta,P}} \T{H^0}
\end{equation}

\noindent By noting that $\T{\varepsilon} << 1$, the Green-Lagrange deformation measure of $\T{F}^\varepsilon$, namely $\T{E}=\frac{1}{2}[(\T{F}^\varepsilon)^{-1} \cdot \T{F}^\varepsilon - \T{I}]$ is well approximated by $\T{\varepsilon}$. Thus, the isothermal spatial elasticity tensor $\T{c}$ is defined through the Taylor expansion in $\T{\varepsilon}$ of Helmholtz free energy $\phi(\T{F^{\theta,P}},\T{\varepsilon}, T)$ 
for the present volume $V$, pressure $P$ and temperature $T \neq T_0$:
\begin{equation}
    \phi(\T{F^{\theta,P}},\T{\varepsilon},T)=\phi(\T{F^{\theta,P}},T) - V(\T{\varepsilon}:\T{I})P + \frac{1}{2}V \T{\varepsilon}:\T{c}:\T{\varepsilon} + ...
\end{equation}

\noindent The $4^\text{th}$order spatial elasticity tensor, defined as $c_{ijkl} = \partial^2\phi/(\partial\varepsilon_{ij} \partial\varepsilon_{kl})$ has, by definition, the major and minor symmetries, with $21$ unique coefficients. The Cauchy stress tensor is then defined as:
\begin{equation}
    \T{\sigma} = (\partial \phi / \partial \T{\varepsilon})/V = \T{c}:\T{\varepsilon} - P\T{I}
\end{equation}

\noindent The elasticity coefficients were calculated for temperatures between 100 and 900 K and for pressures up to 50 GPa by linear regression of stress-strain curves corresponding to Equation 5. The symmetry of stress and strain tensors generally leads to the more convenient \textit{Voigt notation} where a single index replaces pairs of indices. Stress and strain tensors are then expressed in compact notation and read:
\begin{equation}
    \T{\sigma} = [\sigma_{11},\sigma_{22},\sigma_{33},\sigma_{23},\sigma_{13},\sigma_{12}]^T, \quad \T{\varepsilon} = [\varepsilon_{11},\varepsilon_{22},\varepsilon_{33},2\varepsilon_{23},2\varepsilon_{13},2\varepsilon_{12}]^T
\end{equation}
leading to the generalized Hooke's law in tensorial notation:
\begin{equation}
    \sigma_m + \sigma^0_m = c_{mn}^v\varepsilon_n
\end{equation}
where $\T{c}^v$ is referred to as the elasticity matrix (elasticity tensor in Voigt's notations). Each of the 6 independent strains applied have been chosen to be lower than $0.005$ and the reference temperature set to $T_0=300 K$. A total of 25 simulations (6 deformation types times 4 strain amplitudes plus 1 strain-free) is then needed to compute the elastic tensor for one (T, P) state. The elastic tensor has been computed for 5 temperatures and 10 pressures over the range [100, 900 K] and [0, 50 GPa], respectively, requiring a total of 1250 simulations.

\section{Molecular dynamics results}
\subsection{TATB single crystal equation of state}

Following the procedure described in section 2.1, we computed the lattice parameters of TATB single crystal for temperatures between 100 and 900 K and for pressures between 0 and 61.8 GPa. To compare our results against the literature, we present in Figure 2 the 300 K isotherm obtained with the force field used in this work. We compare our results with previous experiments \cite{plisson_equation_2017, olinger_hydrostatic_1976, byrd_ab_2007, stevens_hydrostatic_2008} as well as DFT calculations. 

\begin{figure}[!h]
    \centering
    \includegraphics[height=7cm]{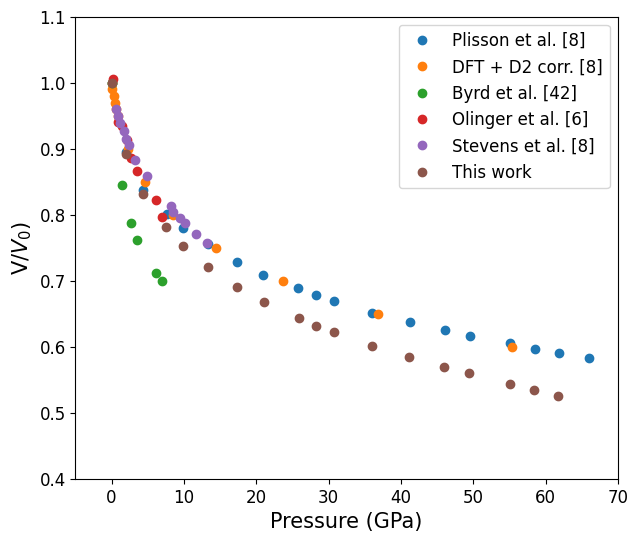}
    \includegraphics[height=7cm]{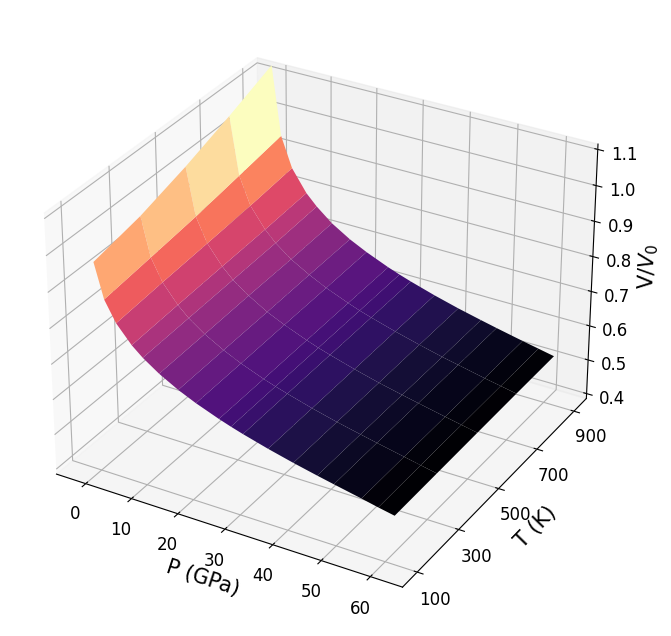}
    \caption{\textbf{(left)} Isothermal EOS of TATB at 300 K. \textbf{(right)} Raw data of TATB EOS obtained through MD simulations.} 
    \label{fig:fig2}
\end{figure}

\noindent The force field used in the present work reproduces the low pressure (P $<$ 10 GPa) experimental data. However, it leads to a larger compressibility at higher pressure that out of the domain where this potential has been fitted (below 13 GPa, after the experimental data from \cite{stevens_hydrostatic_2008}).

A new parametrization of the FF will be the subject of future work, based on the last experimental results as well as reliable DFT calculations that have been recently obtained \cite{plisson_equation_2017, steele_pressure-induced_2019}. Despite this high-pressure bypass, we consider that this FF provides interesting insight to study the evolution of $\T{c}$, and more specifically the evolution of the elastic anisotropy with pressure and temperature. \\

\noindent For the following we take the state (300 K, 0 GPa) as the reference configuration from which the volume ratio is computed as a function of temperature and pressure along isobar and isotherm pathways, respectively. In Figure 3a are represented the five isotherms that have been computed in this work. The volumetric thermal expansion is clearly higher at low pressure than at high pressure where the volume ratio reaches a similar value for the different isotherms, indicating that temperature effects vanish at high pressure. This representation does not allow for a very concise visual analysis so we report in Figure 3b the multiple computed isobars for pressures between 0 and 61.8 GPa. The isobars, instead of isotherms, better show the effect of temperature on the unit cell volume as a function of the hydrostatic pressure. At 300 K, the volumetric thermal expansion is significant with a volume increase of ~15\si{\percent} between 300 K and 900 K. As the hydrostatic pressure increases, the effect of temperature on the volume ratio decreases. The expression of the volumetric thermal expansion is therefore non-linear with pressure and this will be discussed in Section 3 when building analytical laws.

\begin{figure}[!h]
    \centering
    \includegraphics[height=7cm]{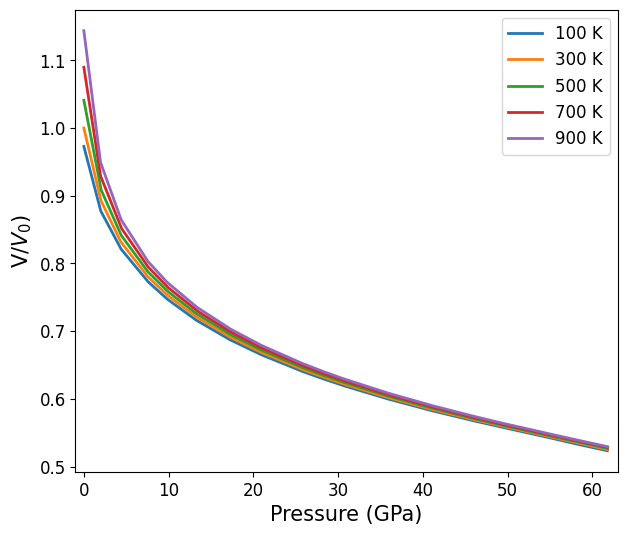}
     \includegraphics[height=7cm]{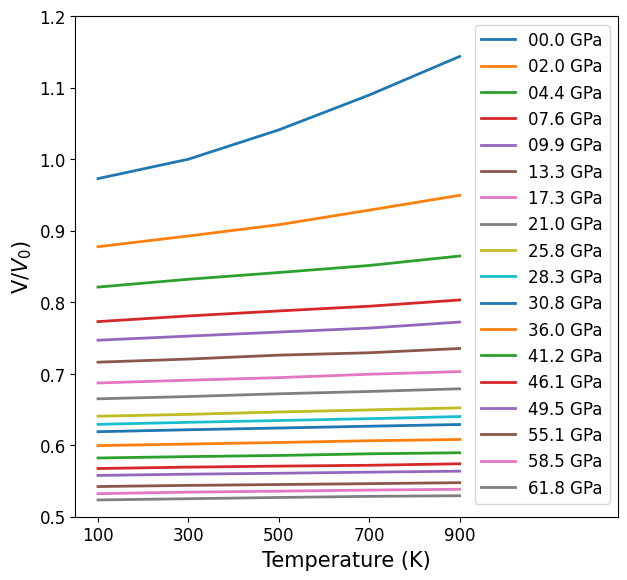}
    \caption{\textbf{(left)} TATB isotherms for T between 100 K and 900 K. \textbf{(right)} Isobars of TATB for P between 0 GPa and 61.8 GPa, where the attenuation of temperature effects is clearly seen at high pressure.}
    \label{fig:fig3}
\end{figure}

\subsection{Isothermal elastic tensor}
\subsubsection{Elastic constants at ambient conditions}
We first discuss our results at ambient conditions, i.e, T = 300 K and P = 0 GPa. The principal components of the elastic tensor computed in this work are compared to literature results \cite{mathew_generalized_2015, lafourcade_dislocation_2017, bedrov_molecular_2009, valenzano_accurate_2012, fan_theoretical_2017, rykounov_investigation_2015, qin_first-principles_2019} in Figure 4. Our predictions are in agreement with the published values, especially with elastic coefficients that were computed with MD simulations under similar conditions. It is found that the predicted values are consistent with other MD results obtained with flexible molecules \cite{mathew_generalized_2015, bedrov_molecular_2009}, whereas the results obtained with rigid molecules seem to overestimate these elastic coefficients \cite{lafourcade_dislocation_2017}. It appears that the elastic coefficients computed by means of ab-initio methods are systematically larger in magnitude than those obtained through classical MD simulations \cite{valenzano_accurate_2012, rykounov_investigation_2015, qin_first-principles_2019}. To this day, no experimental results have been reported for ambient conditions elastic coefficients of TATB single crystal. This is mainly due to the difficulty to synthesize large TATB single crystals. In the following, we present our results for elastic coefficients as function of temperature and pressure but only describe a few isotherms and isobars curves. However, the results obtained for the full range of temperature and pressure are given in the supplementary material. \\

\begin{figure}[!h]
    \centering
    \includegraphics[height=7cm]{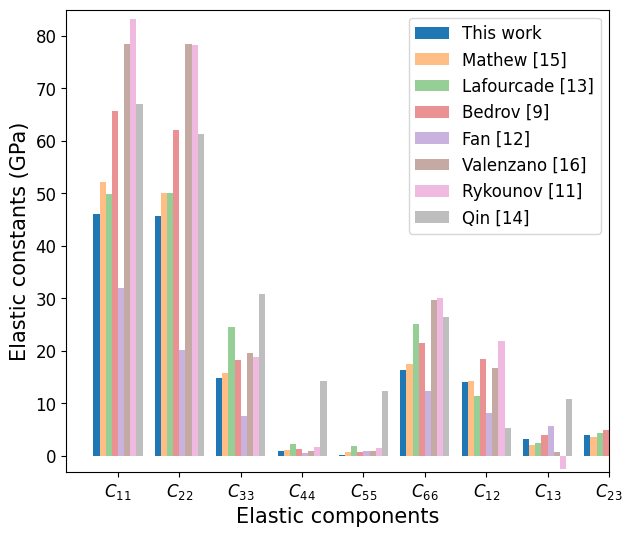}
    \caption{Elastic constants of TATB at ambient conditions, comparison with the literature data.}
    \label{fig:fig4}
\end{figure}

\subsubsection{Effect of hydrostatic pressure at constant temperature}

We then focus on the pressure dependence of TATB elastic constants along two different isothermal pathways to analyze the difference between ambient and high temperatures. The elastic coefficients of TATB along the 300 K and 900 K isotherms are displayed in Figure 5a and 5b, respectively. Concerning the 300 K isotherm, it is clear that all the coefficients are not subjected to the same pressure dependence. Indeed, the coefficients related to the out-of-plane behavior (C33) as well as to the coupling between out-of-plane and in-plane behaviors (C13, C23) seem to depend linearly on pressure. On the contrary, in-plane elastic behavior, mainly governed by constants C11, C22, C66 and C12 exhibits a non-linear evolution with pressure, with a power-law like dependence. These observations correlate with the evolution of lattice parameters of TATB as a function of pressure. Indeed, the inter-layer distance c decreases almost linearly with pressure in contrast with a and b cell parameters which are directly related to the distances between molecules within the layers. This is mainly due to the type of interatomic interactions considered in both cases, namely weak van der Waals interactions for the out-of-plane interactions and strong hydrogen-bonding for the in-plane interactions. \\

\begin{figure}[!h]
    \centering
    \includegraphics[height=6.5cm]{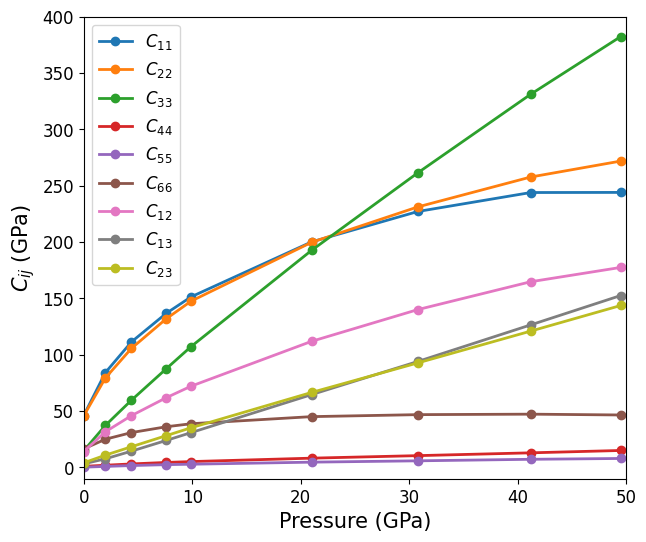}
    \includegraphics[height=6.5cm]{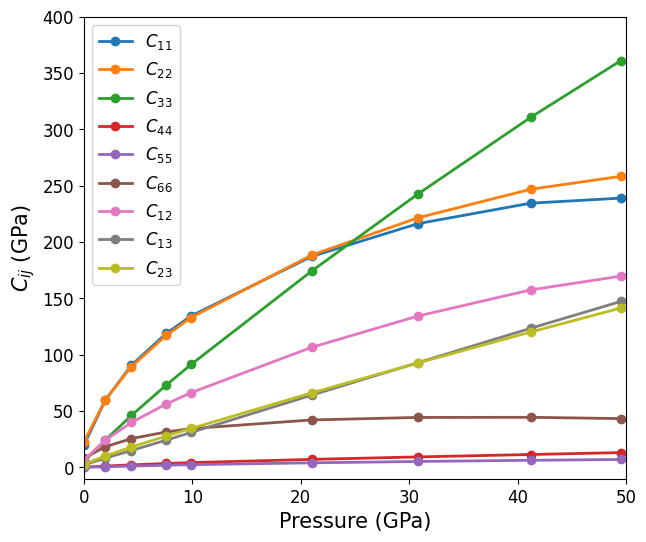}
    \caption{(left) Elastic constants along the 300 K isotherm. (right) Elastic constants along the 900 K isotherm.}
    \label{fig:fig5}
\end{figure}

\noindent We then analyze the 900K isotherm for which the starting configuration is (T=900K, P=0 GPa) and the results are presented in Figure 5b. Besides that TATB single crystal is significantly softer compared to the ambient temperature at 0 GPa, with bulk modulus of 7.71 GPa compared to 16.59 GPa, the same observations as for the 300K isotherm can be made, with a general stiffness slightly lower than at 300 K, along the entire EOS. In addition to these observations, we present in Figure 6a the evolution with pressure of the Voigt (upperscript V) and Reuss (upperscript R) estimates of bulk and shear moduli, denoted $K$ and $G$, respectively. The shear modulus reaches a plateau at high pressure, indicating a saturation in the shear elastic constants as observed in Figure 5. On the opposite, the bulk modulus keeps increasing with pressure in a non-linear fashion. In addition, due to the weak interactions between molecular layers, the shear modulus is uncommonly low, showing that TATB single crystal can be subjected to large shear deformation as observed before with buckling/twinning as well as shear banding deformation mechanisms \cite{lafourcade_irreversible_2018, kroonblawd_high_2020}.\\

\noindent Finally, an assessment of the evolution of anisotropy of single crystal TATB is proposed. When dealing with crystal of complex symmetries, it is useful to refer to the universal anisotropy index \cite{ranganathan_universal_2008}, defined as:
\begin{equation}
    A^U=\T{c}:\T{s} = 5\frac{G^V}{G^R}+\frac{K^V}{K^R}-6 ,
\end{equation}
which takes a value of $0$ for an isotropic material, since Voigt and Reuss estimates for bulk and shear modulus are identical. The universal anisotropy index has been computed along the isotherms and the results are displayed in Figure 6b. At ambient conditions, anisotropy is very high compared to a large panel of materials \cite{ranganathan_universal_2008, kube_elastic_2016} and takes a value of approximately 55 which is among the 25 (over a few thousands available materials) most anisotropic materials if one considers the database of the materials project \cite{jain_commentary_2013}, knowing that most of the highest anisotropic materials are carbon allotropes. Anisotropy strongly decreases with pressure and is divided by almost a factor 10 at 50 GPa compared to 0 GPa. Since elastic anisotropy is the main driver for the buckling elastic instability in TATB (and other materials like graphite \cite{lafourcade_elastic_2020}), such instability would be disadvantaged at high pressure.\\

\begin{figure}[!h]
    \centering
    \includegraphics[height=7cm]{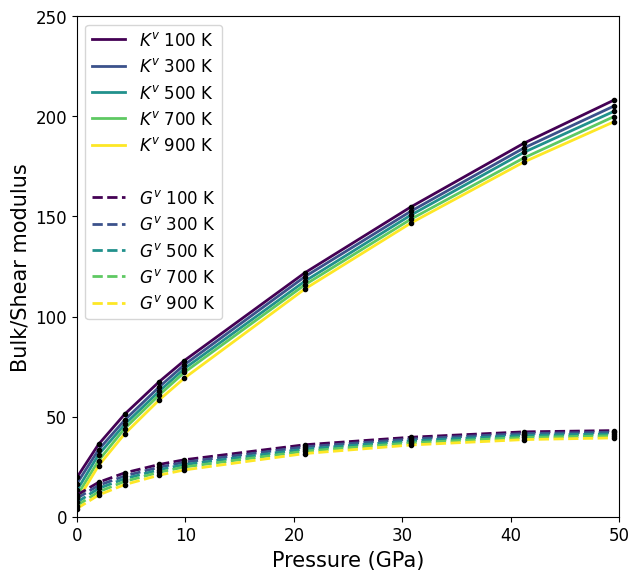}
    \includegraphics[height=7cm]{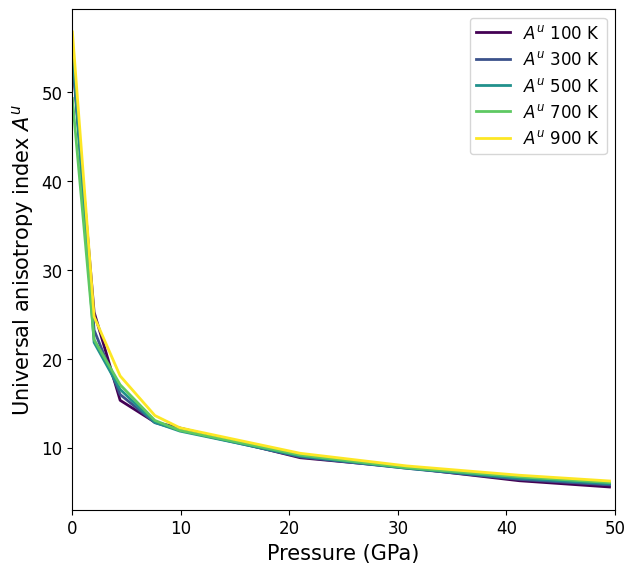}
    \caption{\textbf{(left)} Voigt estimates of bulk and shear modulus as a function of pressure for the different isotherms considered in this work. \textbf{(right)} Corresponding universal anisotropy coefficient as a function of pressure.}
    \label{fig:fig5}
\end{figure}

\noindent In summary, the overall elastic constants dependence with pressure at 900K is similar to 300K. However, one has to consider two groups of elastic constants with different pressure dependence, one being linear and related to out-of-plane elasticity and another one being non-linear and linked to the in-plane stiffness. This will be discussed to propose a closed form expression for P-T dependence of $c_{ij}^v$ in section 4.

\subsubsection{Effect of temperature at constant pressure}
As shown in Figure 7a for ambient pressure, stiffness decreases linearly when temperature increases. However, the slope of this linear dependence is not identical for all elastic constants. Indeed, longitudinal in-plane constants C11 and C22 seem to be more sensitive to temperature than their counterpart C33 which evolves similarly to C12 and C66, both related to the molecular layer elastic behavior under shear and bi-axial stain. Finally, elastic constants related to the coupling between in-plane and out-of-plane behaviors, i.e. C13 and C23, are subjected to a slow decrease in temperature which is also the case for basal plane shear elastic constants C44 and C55. \\

\noindent In Figure 7b are displayed the main elastic constants of TATB single crystal along the 50 GPa isobar in order to analyze the effect of temperature at high pressure. Besides the larger stiffness due to the high pressure conditions, which confirms our previous observations, we can clearly see the difference with the case at 0 GPa. Indeed, the linear dependence of elastic coefficients with temperature is still present, but very less pronounced. Except for longitudinal coefficients, which still exhibit a small decrease with temperature, other coefficients almost remain constant over the entire range of temperature. This attenuation of temperature effects as the pressure increases will be discussed in the next section where analytical laws will be introduced. \\

\begin{figure}[!h]
    \centering
    \includegraphics[height=7cm]{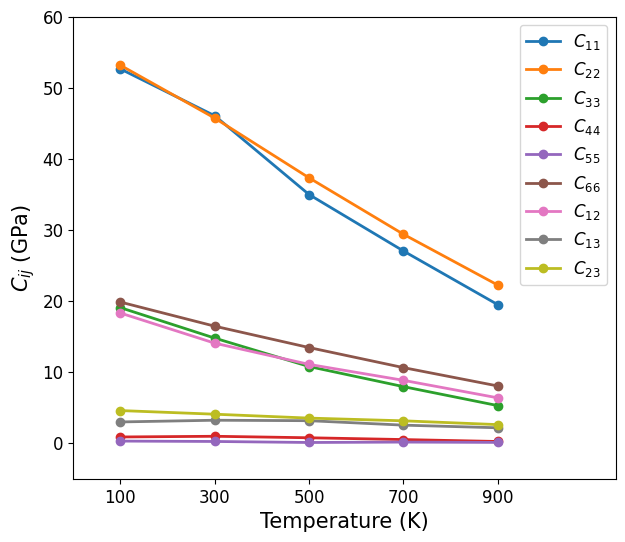}
    \includegraphics[height=7cm]{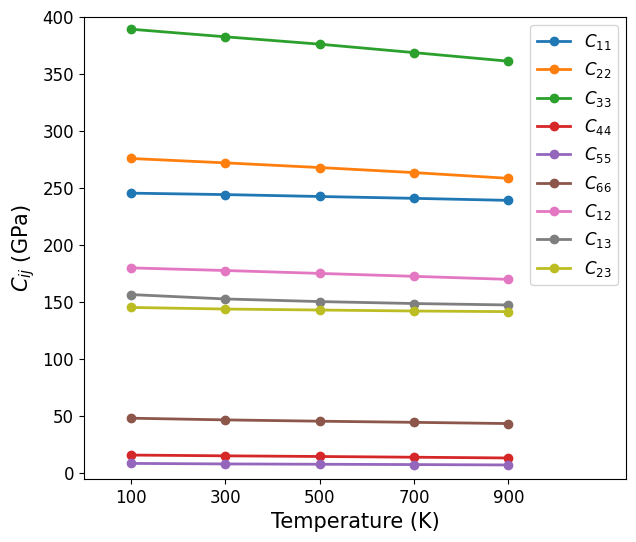}
    \caption{Elastic constants along the 0 GPa \textbf{(left)} and 50 GPa \textbf{(right)} isobars.}
    \label{fig:fig5}
\end{figure}

\noindent In addition to the analysis of single elastic constants, we observe that both bulk and shear moduli evolve linearly with temperature, for pressures between 0 and 50 GPa. Finally, the evolution with temperature of the universal anisotropy index $A^U$ is linear but $A^U$ remains almost constant. This fact has to be considered along with the previous observation that anisotropy strongly decreases with pressure. In particular the buckling instability, controlled by the level of anisotropy, will be impacted by thermodynamic conditions. In the perspective of building analytical laws, we retain the linear dependence of elastic constants with T along the isobars in the regime 0-50GPa.

\section{Temperature and pressure dependent analytical formulations}
\label{sec:analytical}
We propose in the following analytical laws for both equations of state and elastic constants as a function of temperature and pressure. Several EOS models are tested against our data obtained from MD simulations and a polynomial formulation is built for the temperature and pressure dependent elastic coefficients. These laws represent a first step towards a mesoscopic modelling of TATB single crystal that would account for thermal volumetric expansion, high pressure volumetric compression and non-linear elasticity.

\subsection{Equations of state}
\label{subsec:analytical_eos}
Multiple models of EOS were tested against our data, namely Vinet \cite{vinet_universal_1986}, Murnaghan \cite{murnaghan_compressibility_1944} and Birch-Murnaghan \cite{birch_finite_1947} analytical laws. We independently fit the five isothermal curves investigated in this work with the three models and compute the mean absolute error (MAE) as:
\begin{equation}
    MAE^{eos}_{model} = \frac{1}{N}\sum_{i=0}^{N} \Big| P^\mathrm{model}_i - P^\mathrm{MD}_i \Big|
\end{equation}
in order to select the more appropriate model. We plot in Figure 9a the MAE values for the proposed models for the different temperatures. As Vinet and Birch-Murnaghan EOS do not include explicit temperature dependence, we introduce an implicit temperature dependence through the use of different values for the bulk modulus, its derivative and the reference volume as a function of temperature. This is achieved by fitting the parameters by an analytical law independently for each temperature. On the contrary, the Murnaghan EOS already includes a functional form accounting for the thermal volumetric expansion, and its related parameters are then fitted on the whole set of isotherms. 

We plot in Figure 9b the optimized bulk modulus and its pressure-derivative for the three EOS, together with the MD simulation data. It appears that the Vinet EOS leads to the smallest MAE over the different isotherms considered, compared to Birch and Murnaghan models, but also gives more consistent values of $B_0$ and $\partial B / \partial P$ w.r.t MD data. \\

\begin{figure}[!h]
    \centering
    \includegraphics[height=7cm]{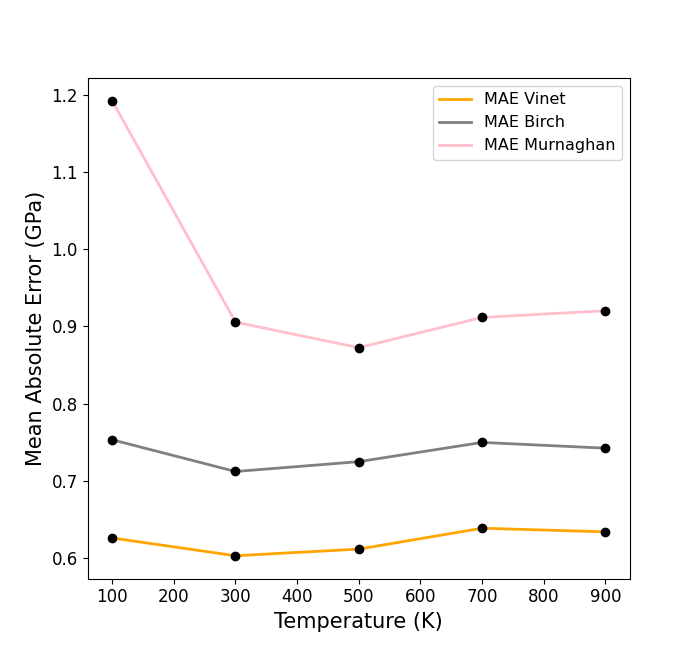}
    \includegraphics[height=7cm]{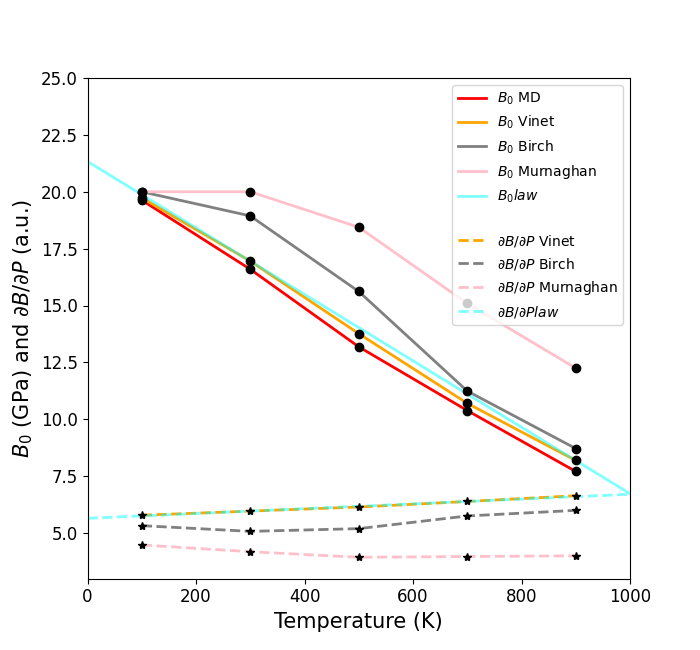}
    \caption{\textbf{(left)} Mean absolute error obtained for the different models on the isothermal data. \textbf{(right)} Comparison between optimized parameters of the different analytical models and the values computed from the MD data.}
    \label{fig:fig5}
\end{figure}

\noindent We then choose to keep the Vinet EOS and make $B_0$ and $\partial B / \partial P$ depend linearly on temperature. Thus, by taking (T=300K, P=0GPa) as the reference state, we have the following expressions:
\begin{equation}
    B_0(T) = B_0^{300K} + \beta \cdot (T-300) 
\end{equation}
\begin{equation}
    B'(T) = B'^{300K} + \beta' \cdot (T-300) 
\end{equation}
\begin{equation}
    V_0(T) = V_0^{300K} + \gamma \cdot (T-300).
\end{equation}
A linear regression on the MD simulation data leads to the values: $\beta= -0.0146039$ \si{\giga\pascal\per\kelvin}, $\beta'= 0.0010561$ \si{\per\kelvin} and $\gamma= 0.1057505$ ang\si{\cubic\angstrom\per\kelvin}. Finally, the temperature and pressure dependent Vinet EOS reads:
\begin{equation}
P(V) = 3 B_0(T) \cdot \frac{1-f(T)}{f^2(T)} \cdot \exp\Bigg(\frac{3(B'(T)-1) \cdot (1-f)}{2}\Bigg)
\end{equation}
where $f(T)=(V/V_0(T))^{1/3}$. The mean absolute error (on pressure) of this equation over the entire volume ratio vs. temperature range spanned in this work is displayed in Figure 9. The general evolution is caught correctly while some drift in pressure appears at high compression ratio. However, we recall that the force field used in this work leads to a higher compressibility for TATB single crystal at high pressure and a more accurate force field would conduct to higher values of elastic constants. However, the procedure described above provide a general framework for choosing the most appropriate EOS model, and do not depends on a particular force field.

\begin{figure}[!h]
    \centering
    \includegraphics[height=7cm]{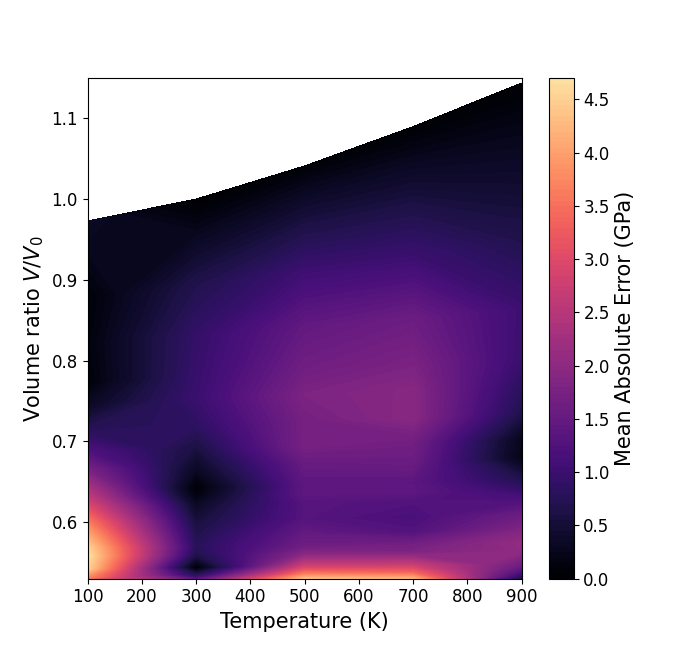}
    \caption{Mean absolute error on pressure, computed over the full range of MD data.}
    \label{fig:fig5}
\end{figure}

\subsection{Elastic constants}
\label{subsec:analytical_cij}
According to our previous observations about the non-linear evolution of elastic coefficients with pressure, we propose a simple polynomial fit of the elastic constants that allows the stiffness to depend on the hydrostatic pressure. Taking the conditions (T=300K, P=0GPa) as the reference state, the isothermal spatial elasticity tensor, along the 300K isotherm, is expressed as:

\begin{equation}
    \TT{c}(P) = \TT{c}^0 \Big(\TT{I} + \sum_{k=0}^{N} \TT{A}_i P^k\Big)
\end{equation}
where $\TT{c}^0$ is the stiffness tensor at the reference state, $\TT{A}^k$ is a tensor containing the polynomial fit coefficients and $N$ is taken equal to 3. Here the multiplication between the reference stiffness tensor and the coefficients tensors is done through the Hadamard product. The pressure-dependent elastic coefficients and their corresponding polynomial fits are represented in Figure 10a. In addition, the polynomial coefficients are provided in the supplementary material.\\

\begin{figure}[!h]
    \centering
    \includegraphics[height=7cm]{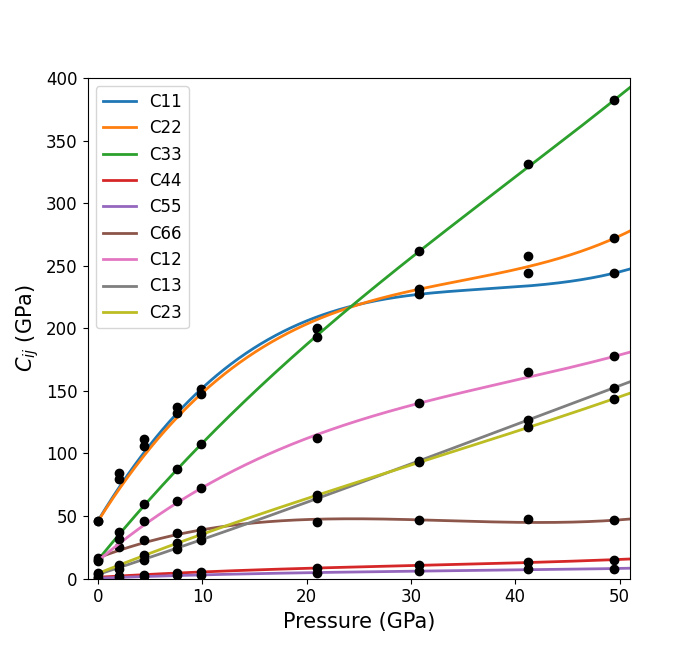}
    \includegraphics[height=7cm]{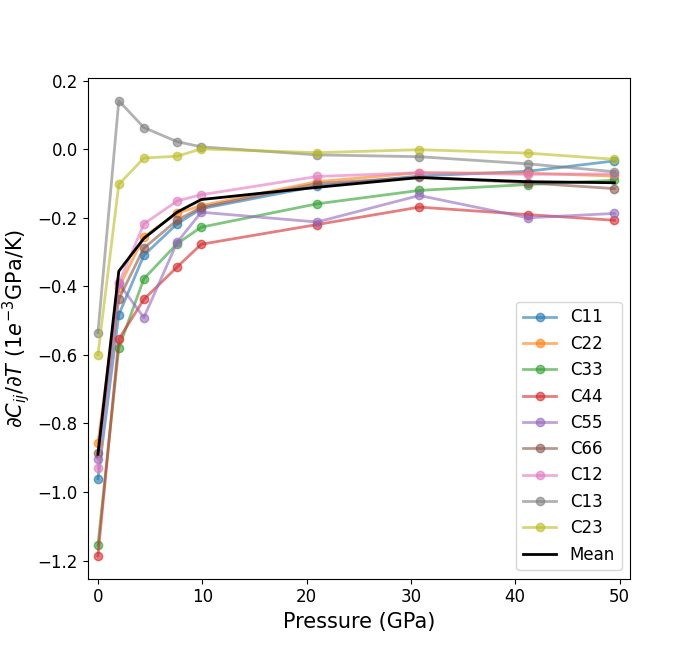}
    \caption{\textbf{(left)} Pressure-dependent elastic moduli of TATB at 300K. The black dots are the MD data and the continuous curves are polynomial fits to the data which are constrained to pass through the reference state values. \textbf{(right)} Partial derivates w.r.t temperature of elastic coefficients at fixed pressure.}
    \label{fig:fig5}
\end{figure}

\noindent The temperature-dependence of elastic coefficients over the entire range of pressure is studied by computing their partial derivatives with respect to temperature at each pressure considered in this work. These partial derivatives as a function of pressure are displayed in Figure 10b. As described previously, elastic coefficients are quite sensitive to temperature at low pressure compared to high pressure; the dark line in figure 10b represents the scalar average partial derivative, denoted $\average{\partial \TT{c}(p) / \partial T}$, which indeed exhibits large variations between 0 and 10 GPa. As a first approximation, one can express the pressure and temperature dependent spatial elasticity tensor as:

\begin{equation}
    \TT{c}(p,T) = \TT{c}^0 \Big(\TT{I} + \sum_{k=0}^{N} \TT{A}_i p^k\Big) \Big( 1 + \average{\partial \TT{c}(p) / \partial T} (T-T_0) \Big)
\end{equation}
where $T_0$=300K is taken as the reference temperature, such that at (T=300k, P=0GPa) the elastic tensor perfectly meets the MD data. This formulation catches the change in anisotropy with pressure through the tensor containing the polynomial coefficients, which is critical for the elastic stability of TATB under pressure. Additionally, the temperature dependence term in this equation maintains, on one hand, the elastic anisotropy along isobaric pathways, and, on the other hand, accounts for the attenuation of elastic softening at high pressure, through the pressure-dependence of the partial derivative term.

\section{Conclusion}
\label{sec:conclusion}
Equilibrium Molecular Dynamics simulations of TATB single crystal have been performed on an extended thermodynamic domain (between 100K and 900K and between 0 and 50 GPa). At each (P,T) point the spatial elasticity tensor has been computed through the derivation of the Cauchy stress with respect to strain. At high pressure (i.e. above the pressure used to build the force field employed in this work), the TATB single crystal appears too compressible. A correction of this force field will be the object of future work. Along isothermal pathways, elastic coefficients are found to increase non-linearly with pressure and elastic anisotropy is subjected to a dramatic decrease, divided by a factor of 3 at 15 GPa and up to 10 at 50 GPa compared to ambient pressure. This could have implications in the ability of TATB single crystal to trigger the twinning-buckling instability, known to be driven by its anisotropy. 

Along the isobaric pathways, increasing temperature leads to a linear decrease of elastic coefficients, while anisotropy remains almost constant. Finally, we introduced preliminary elements to build a mesoscopic model. The first one is a generalized Vinet EOS parametrized over our MD results, which reproduces the volumetric compression as a function of pressure and temperature. This is achieved by making the  bulk modulus, its pressure derivative and ambient pressure reference volume dependent on temperature. The second element is an analytical formulation for the spatial elasticity tensor as a function of pressure and temperature. The dependence on pressure is reproduced through a component-wise polynomial fit while, along isobaric pathways, the temperature dependence is propagated using the average partial derivative of elastic constants with respect to temperature. These elements can be used in mesoscopic models and continuum formulations to model the non-linear elastic behavior of TATB single crystals with pressure as well as its elastic softening in temperature.

\section{Acknowledgments}
The authors would like to thank Dr. Tommy Sewell, Dr Nithin Mathew and Puhan Zhao for their help in building the LAMMPS input decks for TATB flexible molecules simulations. In addition, the authors are thankful for Dr Tommy Sewell and Dr Matthew Kroonblawd for fruitfull discussions about the results obtained through the present force field, as well as for discussing some perspectives concerning its parametrization.
\newpage

\bibliographystyle{ieeetr}
\bibliography{references.bib}
\newpage 

\begin{center}
    \textbf{Supporting Information} \\
    \textbf{Elastic anisotropy of 1,3,5-Triamino-2,4,6-Trinitrobenzene as a function of temperature and pressure: A Molecular Dynamics study}
\end{center}

\section*{TATB elastic constants along isothermal pathways}
Due to the almost transversely isotropic property of TATB single crystal, we present below, the principal components of TATB second-order elastic tensor using Voigt notation, along the different isotherms considered in this work. Tables \ref{tab:cij100K}, \ref{tab:cij300K}, \ref{tab:cij500K}, \ref{tab:cij700K} and \ref{tab:cij900K} show the data for the 100 K, 300 K, 500 K, 700 K and 900 K isotherms, respectively.

\begin{table}[h!]\centering

    \pgfplotstableset{
       every head row/.style={before row=\toprule,after row=\midrule},
       every last row/.style={after row=\bottomrule}
       }
  \begin{filecontents*}{data/cij100K.csv}
P C11 C22 C33 C44 C55 C66 C12 C13 C23
0.000 52.654 53.174 19.071 0.874 0.296 19.854 18.311 2.989 4.589
2.000 92.427 86.253 41.674 2.154 0.905 27.072 34.069 7.682 10.994
4.400 118.924 112.355 65.086 3.500 1.867 32.749 48.637 14.860 18.714
7.600 142.989 137.834 93.041 4.755 2.586 37.598 64.253 24.204 28.483
9.900 156.682 152.611 112.184 5.501 2.917 40.044 74.057 30.946 35.226
21.000 204.050 203.498 198.614 8.611 4.908 45.842 113.672 64.990 66.410
30.800 230.360 234.420 268.192 10.829 5.968 47.466 142.196 94.990 92.922
41.200 246.394 260.970 337.712 13.477 7.543 48.256 167.052 128.596 121.372
49.500 245.338 275.730 389.048 15.747 8.433 48.068 179.824 156.440 145.148
\end{filecontents*}
\pgfplotstabletypeset[
      fixed,fixed zerofill,precision=2,
      column type=r,col sep=space,
      columns={P,C11,C22,C33,C44,C55,C66,C12,C13,C23},
  ]{data/cij100K.csv}
  \caption{TATB principal elastic components along the 100 K isotherm in GPa.}
  \label{tab:cij100K}
\end{table}

\begin{table}[h!]\centering

    \pgfplotstableset{
       every head row/.style={before row=\toprule,after row=\midrule},
       every last row/.style={after row=\bottomrule}
       }
  \begin{filecontents*}{data/cij300K.csv}
P C11 C22 C33 C44 C55 C66 C12 C13 C23
0.000 46.065 45.737 14.783 0.972 0.241 16.460 14.070 3.229 4.078
2.000 84.213 79.426 37.319 1.971 0.920 25.048 31.662 7.658 10.924
4.400 111.410 105.675 59.687 3.132 1.684 30.954 45.998 14.476 18.390
7.600 136.732 131.898 87.519 4.442 2.480 36.104 61.867 23.946 28.207
9.900 151.420 147.596 107.304 5.212 2.869 38.804 72.140 30.936 35.232
21.000 200.164 199.782 192.642 8.248 4.708 45.101 111.914 64.640 66.562
30.800 227.164 231.236 261.508 10.490 5.886 46.869 140.108 94.068 92.798
41.200 243.986 257.748 331.210 12.978 7.246 47.295 164.820 126.526 121.002
49.500 244.082 271.900 382.416 15.069 7.970 46.569 177.530 152.588 143.738
\end{filecontents*}
\pgfplotstabletypeset[
      fixed,fixed zerofill,precision=2,
      column type=r,col sep=space,
      columns={P,C11,C22,C33,C44,C55,C66,C12,C13,C23},
  ]{data/cij300K.csv}
  \caption{TATB principal elastic components along the 300 K isotherm in GPa.}
  \label{tab:cij300K}
\end{table}

\begin{table}[h!]\centering

    \pgfplotstableset{
       every head row/.style={before row=\toprule,after row=\midrule},
       every last row/.style={after row=\bottomrule}
       }
  \begin{filecontents*}{data/cij500K.csv}
P C11 C22 C33 C44 C55 C66 C12 C13 C23
0.000 34.980 37.331 10.784 0.762 0.098 13.454 11.086 3.165 3.529
2.000 76.492 73.244 33.230 1.780 0.895 23.020 29.360 8.046 10.840
4.400 104.831 100.408 55.358 2.860 1.524 29.305 44.094 14.586 18.381
7.600 130.949 126.852 82.865 4.158 2.365 34.672 59.943 23.947 28.129
9.900 146.214 143.011 102.620 4.938 2.788 37.505 70.425 31.111 35.281
21.000 195.758 195.820 186.086 7.887 4.518 44.208 109.932 64.184 66.328
30.800 223.684 228.202 255.366 10.158 5.739 46.144 138.176 93.656 92.850
41.200 240.882 254.524 325.042 12.545 6.969 46.398 162.494 125.446 120.884
49.500 242.410 267.786 375.942 14.491 7.683 45.427 174.986 150.298 142.892
\end{filecontents*}
\pgfplotstabletypeset[
      fixed,fixed zerofill,precision=2,
      column type=r,col sep=space,
      columns={P,C11,C22,C33,C44,C55,C66,C12,C13,C23},
  ]{data/cij500K.csv}
  \caption{TATB principal elastic components along the 500 K isotherm in GPa.}
  \label{tab:cij500K}
\end{table}

\begin{table}[h!]\centering

    \pgfplotstableset{
       every head row/.style={before row=\toprule,after row=\midrule},
       every last row/.style={after row=\bottomrule}
       }
  \begin{filecontents*}{data/cij700K.csv}
P C11 C22 C33 C44 C55 C66 C12 C13 C23
0.000 27.042 29.395 7.956 0.511 0.154 10.632 8.842 2.538 3.147
2.000 67.740 66.424 28.476 1.534 0.776 20.667 26.519 8.053 10.433
4.400 98.371 95.452 51.301 2.613 1.377 27.611 42.344 14.866 18.398
7.600 125.413 122.383 78.489 3.873 2.211 33.216 58.347 24.237 28.164
9.900 141.103 138.636 97.988 4.642 2.658 36.170 68.852 31.384 35.334
21.000 191.650 192.312 180.464 7.539 4.326 43.285 108.344 64.210 66.430
30.800 220.058 224.940 248.876 9.781 5.568 45.336 136.272 93.236 92.660
41.200 237.572 250.566 317.116 12.003 6.664 45.503 159.768 123.968 120.254
49.500 240.796 263.366 368.552 13.857 7.401 44.431 172.426 148.576 142.020
\end{filecontents*}
\pgfplotstabletypeset[
      fixed,fixed zerofill,precision=2,
      column type=r,col sep=space,
      columns={P,C11,C22,C33,C44,C55,C66,C12,C13,C23},
  ]{data/cij700K.csv}
  \caption{TATB principal elastic components along the 700 K isotherm in GPa.}
  \label{tab:cij700K}
\end{table}

\begin{table}[h!]\centering

    \pgfplotstableset{
       every head row/.style={before row=\toprule,after row=\midrule},
       every last row/.style={after row=\bottomrule}
       }
  \begin{filecontents*}{data/cij900K.csv}
P C11 C22 C33 C44 C55 C66 C12 C13 C23
0.000 19.491 22.246 5.287 0.238 0.111 8.051 6.379 2.168 2.612
2.000 59.817 60.186 24.343 1.275 0.604 18.444 24.226 8.307 10.251
4.400 90.724 89.470 46.153 2.309 1.188 25.594 39.980 15.025 18.104
7.600 118.897 117.125 73.056 3.527 1.995 31.477 56.264 24.259 27.862
9.900 134.461 132.999 91.402 4.273 2.455 34.481 66.361 31.062 34.630
21.000 187.218 188.430 174.220 7.160 4.107 42.168 106.570 64.128 66.154
30.800 216.356 221.530 242.638 9.352 5.298 44.386 134.366 92.990 92.718
41.200 234.540 246.888 310.720 11.493 6.376 44.509 157.564 123.540 120.174
49.500 239.010 258.346 361.020 13.194 7.074 43.355 169.762 147.316 141.472
\end{filecontents*}
\pgfplotstabletypeset[
      fixed,fixed zerofill,precision=2,
      column type=r,col sep=space,
      columns={P,C11,C22,C33,C44,C55,C66,C12,C13,C23},
  ]{data/cij900K.csv}
  \caption{TATB principal elastic components along the 900 K isotherm in GPa.}
  \label{tab:cij900K}
\end{table}

\section*{Analytical formulation of temperature and pressure dependent elasticity}
Table \ref{tab:cij_300K_fit} contains the 9 polynomial coefficients of the 3 tensors corresponding to the 3rd order pressure-dependent stiffness along the 300K isotherm.
\begin{table}[h!]\centering

    \pgfplotstableset{
       every head row/.style={before row=\toprule,after row=\midrule},
       every last row/.style={after row=\bottomrule}
       }
    \begin{filecontents*}{data/fit_isotherm_cij.csv}
A A B C
$A^k_{11}$ 3.01668983e-01 -7.80866534e-03  7.00782153e-05
$A^k_{22}$ 2.90516343e-01 -7.30829269e-03  6.98485470e-05
$A^k_{33}$ 6.92044500e-01 -6.59626576e-03  5.58584943e-05
$A^k_{44}$ 5.33743835e-01 -1.05373343e-02  1.14620318e-04
$A^k_{55}$ 1.34282462e+00 -2.69249609e-02  2.60317564e-04
$A^k_{66}$ 1.92645085e-01 -6.22131604e-03  6.21434677e-05
$A^k_{12}$ 5.00098219e-01 -9.15324428e-03  7.65983241e-05
$A^k_{13}$ 8.36111514e-01  3.36861515e-03 -2.79187910e-05
$A^k_{23}$ 8.15928689e-01 -5.28890569e-03  5.62128997e-05
\end{filecontents*}
\pgfplotstabletypeset[
      columns/0/.style={column name=\makecell{component\\ {}},
        string type},
      columns/1/.style={column name=\makecell{$k=1$\\(\si{\per\giga\pascal})\\ {}},
        column type=r,sci,zerofill},
      columns/2/.style={column name=\makecell{$k=2$\\(\si{\per\square\giga\pascal})\\ {}},
        column type=r,sci,zerofill},
      columns/3/.style={column name=\makecell{$k=3$\\(\si{\per\cubic\giga\pascal})\\ {}},
        column type=r,sci,zerofill},
      skip first n=1,
      header=false,
  ]{data/fit_isotherm_cij.csv}
  \caption{Polynomial fit coefficients for the elastic constants along the 300 K isotherm.}
  \label{tab:cij_300K_fit}

\end{table}

Finally, Table \ref{tab:cij_partial_T} contains the values of the mean partial derivatives of TATB elastic tensor w.r.t. temperature along the different isobaric pathways considered in this work.
\begin{table}[h!]\centering

    \pgfplotstableset{
       every head row/.style={before row=\toprule,after row=\midrule},
       every last row/.style={after row=\bottomrule}
       }
    \begin{filecontents*}{data/partial_derivatives.csv}
a b
0.0 -8.89828660e-04
2.0 -3.55752465e-04
4.4 -2.60129417e-04
7.6 -1.83380158e-04
9.9 -1.46889974e-04
21.0 -1.11247624e-04
30.8 -8.27672296e-05
41.2 -9.48606657e-05
49.5 -9.77958704e-05
\end{filecontents*}
\pgfplotstabletypeset[
      columns/0/.style={column name=\makecell{Pressure \\ (\si{\giga\pascal}) \\ {}},
        string type},
      columns/1/.style={column name=\makecell{$<\partial c(P) / \partial T>$ \\ (\si{\giga\pascal\per\kelvin}) \\ {}},
        column type=c,sci,zerofill},
      skip first n=1,
      header=false,
  ]{data/partial_derivatives.csv}
  \caption{Mean partial derivatives of elastic tensor w.r.t. temperature along the different isobaric pathways considered in this work.}
  \label{tab:cij_partial_T}
\end{table}

\end{document}